\documentclass{iopart}

\usepackage{iopams}

\usepackage{euscript,amssymb}

\usepackage{epsfig}

\newcommand{\be}[1]{\begin{equation}\label{#1}}
\newcommand{\ee}{\end{equation}}
\newcommand{\ba}[1]{\begin{eqnarray}\label{#1}}
\newcommand{\ea}{\end{eqnarray}}
\newcommand{\rf}[1]{(\ref{#1})}

\newcommand{\dd}{\dagger}

\def\p{\partial}
\def\a{\alpha}

\def\d{\delta}

\def\G{\Gamma}
\def\sg{\sigma}

\def\vfi{\varphi}

\def\D{\Delta}

\newcommand{\cD}{\mathcal{D}}

\newcommand{\cL}{\mathcal{L}}

\newcommand{\ra}{\rangle}
\newcommand{\la}{\langle}

\def\LLC{(LL^\dd)^*}

\def\wt{\widetilde}

\begin{document}

\title
{Scattering cross section resonance  originating from a spectral
singularity}

\author{Boris F. Samsonov}

\address{Physics Department, Tomsk State
 University, 36 Lenin Avenue, 634050 Tomsk, Russia.}

 \ead{\mailto{samsonov@phys.tsu.ru}}


\begin{abstract}
Using techniques of supersymmetric quantum mechanics,
scattering properties of Hermitian Hamiltonians,
which are related to non-Hermitian
ones by similarity transformations, are studied.
We have found that the scattering matrix of the Hermitian Hamiltonian
coincides with the
 phase factor of the non-unitary
scattering matrix of the non-Hermitian Hamiltonian.
The
possible presence of a spectral singularity in a non-Hermitian
Hamiltonian translates into a pronounced resonance
in the scattering cross section of
its Hermitian counterpart.
This opens a way for detecting spectral singularities in scattering
experiments;
although a singular point is inaccessible for
the Hermitian Hamiltonian, the Hamiltonian
 ``feels'' the presence of the singularity if it is ``close enough''.
We also show that cross sections of the non-Hermitian Hamiltonian do not
exhibit any resonance behavior and explain the resonance behavior of the
Hermitian Hamiltonian cross section by the fact that corresponding
scattering matrix, up to a background scattering matrix, is a square root of
the Breit-Wigner scattering matrix.
\end{abstract}


\section{Introduction}

Since the pioneering paper of Bender and Boettcher \cite{Bender-B}, non-Hermitian
Hamiltonians
with a purely real spectrum
have attracted more and more attention from both  theoretical and
 experimental viewpoints.
Many interesting properties of these Hamiltonians have been described in the
literature
 (for a review see \cite{Bender-review}).
If a non-Hermitian Hamiltonian with a real spectrum can never be
 equivalent to a Hermitian one, probably it exhibits the most unusual
 properties.
 These are the Hamiltonians that possess
 exceptional points in the discrete part of the spectrum
 or that have spectral singularities in its continuous
 part.
 In particular,
the property of a super fast evolution,
 called
  ``faster than Hermitian evolution", has been discovered \cite{Brach}.
This was
 explained by noting the presence of an exceptional point in the spectrum of
 a matrix Hamiltonian
 \cite{GS}
 and analyzed from the point of view of
 the Naimark extension \cite{GS-PRL}.
 It has recently been shown that the presence of a spectral singularity leads
 to infinite reflection and transmission coefficients at a real energy
 of a complex scattering potential and
 it has been conjectured that this
 effect is similar to a zero width resonance
 that may be observed in resonating
waveguides \cite{Most-PRL-PRA2009}.
A similar effect was described for a
non-Hermitian Friedrichs-Fano-Anderson model describing the decay of a
discrete state coupled to a continuum of modes and in a Bragg scattering
in complex crystals \cite{Longhi-PRA-2009-20010}.

As far as we know, these unusual properties
are mainly related with
non-Hermitian Hamiltonians.
Nevertheless,
despite the fact that these Hamiltonians have purely real spectra,
 from the point of view of the standard
 quantum mechanics
  they may not be fundamental since they cannot be
 directly associated with quantum mechanical observables.
For this reason
there is no known way to use quantum mechanical experiments to observe
these properties.
On the other hand, any non-Hermitian
diagonalizable  Hamiltonian $H$
with a real and discrete spectrum
possesses a
Hermitian counterpart $h=h^\dag$ that is related to $H$
by a similarity transformation.

For simplicity we will consider the case when both $H$ and $h$ have purely
continuous spectra
and depend on a complex parameter $a$,
$H=H(a)$, $h=h(a)$.
At $a=a_0$ the Hamiltonian $H(a_0)$ has a spectral singularity.
In this case $H(a)$ has no exceptional points, and we
will analyze the scattering properties of $h(a)$ for $a\approx a_0$.
We study this case because $h(a)$, $a\ne a_0$,
 being a usual quantum mechanical Hamiltonian,
exhibits
a number of unusual properties.
These properties have not been described in the literature so far because up to
now no systematic approach for finding the similarity
transformation in an explicit form exists.
In the majority of cases the transformation can only be calculated
using an approximation scheme, notably
perturbation theory \cite{Jones}.
On the other hand, as shown in \cite{my,my-SS}, supersymmetric quantum
mechanics (SUSY QM) may be extremely
useful for studying different properties of
non-Hermitian Hamiltonians.
In the present paper
we will
use the techniques of SUSY QM for constructing the family $H(a)$.
This permits us
for $a\ne a_0$ to express the similarity transformation
from  $H(a)$ to  $h(a)$ in terms
of {\it SUSY transformation operator} (see below)
and to calculate the scattering matrix for both $H(a)$ and
$h(a)$.
 We shall show that the scattering matrix of $h(a)$ coincides with the
phase factor of the scattering matrix of $H(a)$
and for $a\approx a_0$ the scattering cross section for
$h(a)$ exhibits a pronounced resonance.
 Moreover, we shall also show that up to a background scattering matrix,
 the scattering matrix of $h(a)$ is a square root of the Breit-Wigner
 scattering matrix.
 Note that the square root form of the scattering matrix
 appears in some models in atomic \cite{Stewart}
  an nuclear physics \cite{Durand}.
Finally we illustrate our findings
using the simplest but realistic toy model.


\section{Real scattering potentials}

It is well known (see e.g. \cite{Landau}) that
if the scattering particles interact by a potential that depends on the
magnitude of the distance between the particles, the scattering problem
is reduced to solving the radial Schr\"odinger equation
with the so called scattering (asymptotic) condition at infinity.
Therefore
we start with a Hermitian scattering Hamiltonian
$h_0$ defined by a real valued potential $v_0(x)=v_0^*(x)$
\[
h_0=h_0^\dd=-\frac{d^2}{dx^2}+v_0(x)\,,\ x\in [0,\infty)
\]
on a proper domain  $\cD_{h_0}$
in the space $\cL^2$ of functions that are square
integrable on the positive semiaxis.
We will assume that the potential $v_0$ corresponds to
the zero value of the angular momentum $\ell=0$
in the three dimensional scattering process.
In general $v_0(x)$ may be
not only finite at the origin but also
singular.
In the usual radial Schr\"dinger equation this singularity
corresponds to the centrifugal term $\ell(\ell+1)x^{-2}$. Nevertheless,
while modeling real scattering experiments, in many cases one needs
potentials with an $\ell$-independent singularity at $x=0$
(for a discussion see e.g. \cite{ourPRC2002}).
For this reason we will assume that
\[
v_0(x)\to\nu(\nu+1)x^{-2}\,,\quad x\to0
\]
where the parameter $\nu=0,1,2,\ldots$ is
called the singularity strength (see, e.g., \cite{ourPRC2002}).
The functions $\psi\in\cD_{h_0}$ are assumed to be
smooth enough and to
 satisfy the Dirichlet boundary condition at the
origin, $\psi(0)=0$.
For simplicity, we also assume that $h_0$ has a purely continuous
and non-negative spectrum $E=k^2\ge0$
and that
 the eigenfunctions $\psi_k(x)$
\be{Hpsi}
h_0\psi_k=k^2\psi_k\,,\ \psi_k(0)=0\,,\ k\ge0
\ee
satisfy the asymptotic condition at
infinity (see, e.g., \cite{Landau}, Ch. XVII)
\be{psikJ}
\psi_k(x)\propto\psi_{ka}=
A(k)e^{ikx}+B(k)e^{-ikx}\,, \quad x\to\infty\,.
\ee
From here one finds the $S$-matrix and phase shift $\d$
\be{SE}
S=S(k)=\exp\left[2i\d(k)\right]=-\frac{A(k)}{B(k)}
\ee
which define the scattering amplitude
\be{fk}
f(k)=\frac{1}{2ik}\bigl(S(k)-1\bigr)\,,
\ee
effective range function
\be{gk}
g(k)=k\cot\d(k)=ik\,\frac{S(k)+1}{S(k)-1}\,,
\ee
and cross section
\be{sg}
\sg(k)=4\pi|f(k)|^2=\frac{\pi}{k^2}\bigl|S(k)-1\bigr|^2\,.
\ee
We assume here that
all these quantities correspond to zero value of the orbital
momentum, i.e., to $s$-wave which may dominate other partial waves
in the case of a low energy scattering \cite{Landau}.

For the Hermitian Hamiltonian $h_0$,
the $S$-matrix is unitary, $|S(k)|=1$, and
the functions $\psi_k$ form an orthonormal (in the sense of distributions)
and complete basis in the space $\cL^2$
(see, e.g., \cite{Levitan}),
\be{ort-complet-psik}
\la\psi_k|\psi_{k'}\ra=\d(k-k')\,,\quad
\int_0^\infty\!\!dk\, |\psi_k\ra\la\psi_k|=\textbf{1}\,.
\ee

Below we will find asymptotic behavior of the scattering state for a
complex SUSY partner $H$ of the Hamiltonian $h_0$. Therefore the
coefficients $A(k)$, $B(k)$, scattering matrix $S(k)$,
 phase shift and cross section
for the Hamiltonian $h_0$ will be denoted as $A_0(k)$, $B_0(k)$,
$S_0(k)$, $\d_0(k)$ and $\sg_0$ respectively.

\section{Complex SUSY partners of real scattering potentials}

Assuming that we know a solution $u(x)$ of the differential equation
\be{eqa}
h_0u(x)=\a u(x)\,,\quad \a\in\mathbb C\,,
\ee
 we can construct
an exactly solvable non-Hermitian
Hamiltonian
\[
H=-\frac{d^2}{dx^2}+V(x)\,,\quad V(x)\ne V^*(x)
\]
with a real spectrum
using the approach of SUSY QM \cite{Andr-real}.
In this approach,
operators $h_0$, $H$
and $H^\dd=-\p_x^2+V^*(x)$ are related via
intertwining relations
\be{inter}
Lh_0=HL\,,\quad h_0L^\dd=L^\dd H^\dd
\ee
where $L$ is assumed to be a differential operator.
In the simplest case, that we will consider below, operator $L$ is a
first order differential operator
\be{L}
L=-\frac{d}{dx}+w\,,\ w=w(x):=[\log u(x)]'\,.
\ee
Here and in what follows the prime denotes the derivative with respect to
$x$.
The function $w(x)$ is known as a superpotential
(complex valued in the current case)
defined with the help of a complex valued solution
$u=u(x)\ne0$
$\forall x\in(0,\infty)$ to the
differential equation \rf{eqa}
where $\alpha$ is
a complex factorization constant.
The constant $\alpha$ participates at the factorization of the Hamiltonians
$h_0$ and $H$,
\be{factor}
h_0={(L^*)}^\dd L+\alpha\,,\
H=L{(L^*)}^\dd+\alpha\,.
\ee
The potential $V(x)$ is expressed via the superpotential in the usual way,
\be{Vsusy}
V(x)=v_0(x)-2w'(x)\,.
\ee

Any solution $\vfi_k$ of the differential equation
\be{Hphi}
H\vfi_k=k^2\vfi_k\,,\ k\in\mathbb C\,,\ k^2\ne\a
\ee
may be obtained by applying the operator $L$ to a solution to Eq.
\rf{Hpsi},
\be{fikL}
\vfi_k=L\psi_k\,.
\ee
For $k^2=\a$, a solution to this equation is the function
$\vfi_{\sqrt{\a}}=1/u$
(see e.g \cite{Andr-real,my-prepr}).
Another solution to Eq. \rf{Hphi}
corresponding to $k^2=\a$, $\wt\vfi_{\sqrt{\a}}(x)$,
which is linearly independent with $\vfi_{\sqrt{\a}}(x)$,
may be obtained from the
property
\[
 \vfi_{\sqrt{\a}}(x){\wt\vfi}_{\sqrt{\a}}^{\,\prime}(x)-
 \vfi_{\sqrt{\a}}^{\,\prime}(x)\wt\vfi_{\sqrt{\a}}(x)=1\,.
 \]
To select ``physical'' solutions to Eq. \rf{Hphi}, we will solve it with
the Dirichlet boundary condition at the origin, $\vfi_k(0)=0$,
and the asymptotic condition at infinity \rf{psikJ}.

In general, the
transformation operator $L$
violates boundary conditions.
The spectral problem for the Hamiltonian $H$ is simplified
if $L$ transforms eigenfunctions of the Hamiltonian $h_0$ to
eigenfunctions of the Hamiltonian $H$.
Such transformations are known in the literature as conservative
\cite{our}.
For simplicity, we will consider below only conservative transformations.
We thus assume that $L$ preserves the Dirichlet
boundary condition at the origin.
In this case all spectral points of the Hamiltonian $h_0$
(note that they are real!) are also
the spectral points of the Hamiltonian $H$.
 Therefore the Hamiltonian $H$ has a
real spectrum provided the complex factorization constant $\a$ does not
belong to its spectrum.

Since for $k=\sqrt{\a}$ the function $\vfi_{\sqrt{\a}}=1/u$ is a solution to Eq.
\rf{Hphi},
to ensure that
 the spectrum of the operator $H$ is real for a complex $\a$, we
choose the function $u(x)$ such that
\be{fisqrt}
\vfi_{\sqrt{\a}}(x)=1/u(x)\to\exp(-a x)\to\infty\,,\quad x\to\infty
\ee
with
\[
\a=-a^2\,,\quad a=d+ib\,,\quad d<0\,.
\]
We note that the function $u(x)$
with the asymptotic behavior as given in Eq. \rf{fisqrt} is the
 Jost solution to Eq. \rf{eqa}.
(For a more detailed discussion see \cite{my-prepr}.)
Below we will need the asymptotic form $L_a$ as $x\to\infty$
of the transformation operator
\rf{L} which follows from \rf{fisqrt}
\be{La}
L_a=-\p_x+a=-\p_x+d+i b\,.
\ee

We emphasize that for the first order transformation  \rf{L}
to preserve the Dirichlet boundary condition at the origin,
the initial potential should be singular at the origin.
In this case, the SUSY transformation decreases the singularity
strength $\nu$ by one unit
and at $\nu=0$ the potential becomes finite at $x=0$
(see, e.g., \cite{ourPRC2002}).
As it was shown in \cite{my-prepr}
 for $d<0$ the functions
\be{phik}
\vfi_k=(k^2-\a)^{-1/2}L\psi_k\,,\quad k\ge0
\ee
form a continuous biorthonormal set in $\cL^2$.

Let us denote $H|_{d=0}=H_0$.
According to Eq. \rf{fisqrt}, if $d=0$,
the eigenfunction of $H=H_0$ at $k^2=k_0^2=\a=b^2$,
$\vfi_{k_0}(x)=1/u(x)$,
has a pure exponential asymptotic behavior at infinity,
\[
\vfi_{k_0}(x)\to\exp(-ibx)\,,\quad x\to\infty\,.
\]
We emphasize that due to
the choice of the transformation function, the function
$\vfi_{k_0}(x)$ satisfies both the Dirichlet boundary condition at the
origin, $\vfi_{k_0}(0)=0$,
and the asymptotic condition at infinity \rf{psikJ}
with
either $A(k)=0$ if $b>0$ or $B(k)=0$ if $b<0$.
For these reasons, the point $k^2=k_0^2=\a=b^2$ belongs to the
continuous spectrum of $H_0$ but
the resolution of the identity operator over the set of the eigenfunctions of
$H_0$ becomes divergent at $k=k_0$ \cite{Liantse} and needs a special
regularization procedure \cite{Andr,my-prepr}.
The spectral point
$E=k_0^2$
possessing such properties
is known in the literature as the spectral singularity.
(The interested reader can find a rigorous mathematical
definition of the spectral singularity in \cite{Schw},
see also a recent discussion in
\cite{Most-PRL-PRA2009,Andr,my-prepr} and references in that papers.)
Moreover, if $H$ possesses a spectral singularity, there is no way to
relate it to a Hermitian Hamiltonian by a similarity transformation.
Therefore, below we will assume $d<0$
since just in this case the functions \rf{phik}
form a continuous biorthonormal set in $\cL^2$ \cite{my-prepr}
and will analyze properties
of a Hermitian counterpart of $H$ when $d\to-0$ and $H\to H_0$.


\section{Hermitian operator equivalent to $H$}

To establish an equivalence between the non-Hermitian operator $H$ and a
Hermitian operator $h$ we will use ideas formulated in \cite{Scholtz}
for quasi-Hermitian Hamiltonians
and further developed in \cite{Most_JPA} for pseudo-Hermitian
Hamiltonians combined with the supersymmetric approach. In this way we
will express the equivalence transformation in terms of the SUSY
transformation operators.

From \rf{inter} one easily deduces that
\be{LLH}
\eta H=H^\dd\eta\,, \quad \eta:=\LLC\,.
\ee
The operator $\eta$, as introduced in \rf{LLH} with
$L$ given in \rf{L},
is
 positive definite, Hermitian and invertible \cite{my-prepr}
 second order differential operator
\[
\eta=(-\p_x+w^*)(\p_x+w)=\eta^\dagger\,.
\]
It has a unique Hermitian, positive definite
and invertible square root
\be{rho}
\rho=\eta^{1/2}=\rho^\dag\,,\quad \rho>0\,.
\ee
Hence, from Eq. \rf{LLH} one finds
the Hermitian operator $h$ equivalent to $H$,
\be{h}
h=\rho H\rho^{-1}=\rho^{-1}H^\dd\rho=h^\dag
\ee
which also has a purely continuous
real and non-negative spectrum.
Its eigenfunctions $\Phi_k$,
\[
h\Phi_k=k^2\Phi_k\,,\quad k\ge0\,,
\]
 are obtained by applying the operator $\rho$
to the eigenfunctions of $H$
\be{Phik}
\Phi_k=(k^2-\a)^{-1/2}\rho\,\vfi_k=(k^2-\a)^{-1}\rho L\psi_k\,.
\ee
The factor $(k^2-\a)^{-1/2}$ is introduced to guarantee both the normalization of
these functions,
$\la\Phi_{k'}|\Phi_k\ra=\d(k-k')$,
and their completeness \cite{my-prepr}.
From Eq. \rf{Phik} it follows that the eigenfunctions $\vfi_k$ of $h_0$ and
 $\Phi_k$ of $h$ are related by an isometry $\mathbf{U}$
\be{UPhik}
\Phi_k=\mathbf{U}\psi_k
\ee
where
\[
\mathbf U=\rho L(h_0-\a)^{-1}=L^*[(L^\dag L)^*]^{-1/2}=
\bigl(\mathbf{U}^\dag\bigr)^{-1}\,.
\]
Another remarkable property
of the operators $\rho$ and $\rho^*$
 is that their product factorizes a polynomial of the
operator $h$
\[
\rho(\rho^*)^2\rho=(h-\a)(h-\a^*)\,,
\]
which follows from Eqs. \rf{inter},
\rf{factor} and \rf{h}.
Using the spectral decomposition of the operator $h$,
 one can express $h$ in terms of $\rho$, $L$, $L^\dd$ and the
resolvent of $h_0$,
\be{hrho}
h=\frac{\rho L}{\a-\a^*}\,
\left[
\a(h_0-\a)^{-1}-\a^*(h_0-\a^*)^{-1}
\right]
L^\dd\rho\,.
\ee
Note that for any $a=d+ib$ with $d<0$ the point $\a=-a^2$ does not
belong to the spectrum of $h_0$ and the operator \rf{hrho} is well
defined.

Although Eq. \rf{Phik}  [or \rf{UPhik}]
 formally solves the problem of finding the eigenfunctions of
$h$, it contains the non-local operator $\rho$ (or $[(L^\dag
L)^*]^{-1/2}$)
and, therefore, in general, no explicit expression for $\Phi_k$ exists.
Fortunately, to describe the scattering properties of $h$, one needs
 only the asymptotic form of these functions.


\section{Scattering matrices and cross sections for $H$ and $h$}

According to Eqs. \rf{psikJ} and \rf{SE},
the $S$-matrix for a Hermitian Hamiltonian is
defined by the asymptotic behavior
(i.e., as $x\to\infty$) of the corresponding
scattering state.
 In case of an inelastic scattering, the non-unitary scattering matrix is
 defined by the same Eq. \rf{SE} \cite{Landau}
where  $A(k)$ and $B(k)$ should be replaced by
 $A_H(k)$ and $B_H(k)$ found from the asymptotic behavior of
the function $\vfi_k$ given in \rf{fikL},
  \rf{psikJ} and \rf{SE} \cite{Landau}.
Now using Eqs. \rf{phik} and \rf{La}, we express the
asymptotic form $\vfi_{ka}$ of the function $\vfi_k$
in terms of the Hamiltonian $h_0$ scattering states asymptotics
$\psi_{ka}$,
\be{fikas}
\vfi_{ka}=L_a\psi_{ka}=(-\p_x+d+ib)\psi_{ka}\,.
\ee
Here we dropped the inessential normalization factor $(k^2-\a)^{-1/2}$,
so that $\vfi_k\propto\vfi_{ka}$ as $x\to\infty$.

Using Eq. \rf{psikJ}, where coefficients $A(k)$ and $B(k)$ are replaced by
$A_0(k)$ and $B_0(k)$ respectively, we get for $\vfi_{ka}$ the same
expression \rf{psikJ} with the coefficients
\ba{AH}
A(k)\equiv A_H(k)=A_0(k)(d+ib-ik)\,,\\
B(k)\equiv B_H(k)=B_0(k)(d+ib+ik)\,.\label{BH}
\ea
From here
it follows the scattering matrix $S_H$ for the Hamiltonian $H$
\be{SH}
S_H(k)=-\frac{A(k)}{B(k)}=S_0(k)\wt S(k)
\ee
where
\[
\wt S(k)=\frac{d+ib-ik}{d+ib+ik}\,.
\]
We note a non-unitary character of $S_H$,
\be{absS}
|S_H(k)|=|\wt S(k)|=\left[\frac{(b-k)^2+d^2}{(b+k)^2+d^2}\right]^{1/2}.
\ee
As is known from the theory of inelastic collisions \cite{Landau},
 $S_H$ defines the elastic cross section
$\sg_e=\frac{\pi}{k^2}|S_H-1|^2$.
Its absolute value $|S_H|$ determines
a reaction cross section, $\sg_r=\frac{\pi}{k^2}(1-|S_H|^2)$.
The total cross section is their sum, $\sg_t=\sg_e+\sg_r$.

Using Eq. \rf{Phik} and dropping once again the factor $(k^2-\a)^{-1/2}$,
 we find the  asymptotic form
 $\Phi_{ka}$ of
the function $\Phi_k$,
\[
\Phi_{ka}=\rho_a\vfi_{ka}
\]
where $\rho_a$ is the asymptotic form of the operator
$\rho$.

To find the action of the operator $\rho_a=\eta_a^{1/2}$
on the function \rf{fikas},
we notice that according to \rf{LLH} and \rf{La}
\[
\eta_a=(L_aL_a^\dag)^*=(-\p_x+d-ib)(\p_x+d+ib)
\]
and
therefore that
\[
\eta_ae^{ikx}=[(k+b)^2+d^2]e^{ikx}\,.
\]
From this, one finds the action of the operator
$\rho_a=\eta_a^{1/2}$ on the exponential
function
\be{rhoae}
\rho_ae^{ikx}=\sqrt{(k+b)^2+d^2}e^{ikx}\,,
\ee
which finally yields the asymptotic form \rf{psikJ} of the
Hamiltonian $h$ scattering state $\Phi_{ka}$
with the coefficients
\be{Ak}
A(k)=A_H(k)\,[\,(k+b)^2+d^2\,]^{1/2}
\ee
and
\be{Bk}
B(k)=B_H(k)\,[\,(k-b)^2+d^2\,]^{1/2}\,.
\ee
We would like to stress the presence of the square root branch points in
these formulas which come from Eq. \rf{rhoae}.
Here, in agreement with Eq. \rf{rho},
 only one branch of the square root is chosen.

Using Eqs. \rf{AH} and \rf{BH} one sees that the coefficients
 \rf{Ak} and \rf{Bk} contain the common factor
 $[\,(d+ib)^2+k^2\,]^{1/2}$ which will cancel out in the ratio
$A(k)/B(k)$.
By this reason, the asymptotics of the state $\Phi_k$ may be
written in the form \rf{psikJ} with the coefficients
\ba{Ah}
A(k)\equiv A_h(k)=A_0(k)[\,(d-ik)^2+b^2\,]^{1/2}\,\hphantom{.}\\
B(k)\equiv B_h(k)=B_0(k)[\,(d+ik)^2+b^2\,]^{1/2}\,.
\ea


Once the asymptotic form of the scattering state is established, one finds
 the
scattering matrix $S_h$ for $h$ with the help of the asymptotic boundary
condition \rf{psikJ}
\be{Sh}
S_h(k)=S_0(k)S_R(k),\hphantom{i}
S_R(k)=\wt S(k)
\left[
\frac{(b+k)^2+d^2}{(b-k)^2+d^2}
\right]^{1/2}\hspace{-.5em} .
\ee
Comparing equations \rf{SH}, \rf{absS} and \rf{Sh}, we see that
the phase factor of $S_H$ coincides with the scattering matrix
 of the Hermitian Hamiltonian $h$ equivalent to $H$,
 \be{SHhH}
S_h=\frac{S_H}{|S_H|}\,.
\ee

We note that the scattering matrix
\be{S2R}
S_{BW}=S_R^{\,2}(k)=\frac{b^2+(d-ik)^2}{b^2+(d+ik)^2}
\ee
leads to
a Breit-Wigner resonance formula (see e.g. \cite{Perkins})
\[
\sigma_{BW}=\frac{16\pi d^2}{(k^2+d^2-b^2)^2+4b^2d^2}
\]
which in the energy scale  reads
\be{sgBW}
\sigma_{BW}=\frac{4 \pi}{b^2}\,\frac{(\G/2)^2}{(E-E_0)^2+(\G/2)^2}
\ee
with $\G=4 bd$ and $E_0=b^2-d^2$.
We assume $|d|$ is small enough so that $b^2>d^2$.
Near the resonance $E\approx E_0$ and $k\approx b$ so that Eq. \rf{sgBW}
 reduces to the celebrated Breit-Wigner formula
 (see e.g. \cite{Bohm}).
 From here we conclude that the $S$-matrix $S_R$ is a square root of the
 Breit-Wigner $S$-matrix $S_{BW}$ given in \rf{S2R}.
  Note that an electron scattering $S$-matrix square root
 plays an essential role in the theory of
 spin effects in photoionization developed by Stewart \cite{Stewart} who
 noticed that ``a photoionization experiment contains half of an electron
 scattering experiment''.
A square root of a unitary $S$-matrix appears also in some models
describing absorptive processes in high energy reactions
involving elementary particles \cite{Durand}.

The phase shift $\d_R$ corresponding to $S_R$ is one half of $\d_{BW}$,
$\d_R=\frac12\d_{BW}$.
It leads to a cross section with a square root branch point
\be{sgR}
\sg_R(k)=\frac{2\pi}{k^2}\left[
1+\frac{k^2-b^2-d^2}{\sqrt{(k^2+d^2-b^2)^2+4b^2d^2}}
\right].
\ee
As it was already mentioned, we choose here that sign of the square root
which corresponds to positive definite operator $\rho$ \rf{rho}.
It is not difficult to see that
$\sg_R(0)=4\pi d^2/(b^2+d^2)^2>0$, $\lim_{k\to\infty}\sg_R(k)=0$,
$\sg_R(k)>0$ for $0\le k<\infty$ and $d\sg_R(0)/dk>0$ for
$b^2>\frac12d^2$.
These results mean that for any fixed value of $b$ and small enough value
of $|d|$, the function $\sg_R(k)$ \rf{sgR} has a maximum and therefore exhibits
resonance behavior.
This evidently  is just a consequence of the fact that
the scattering matrix $S_h$ \rf{Sh} is a square root of $S_{BW}$ \rf{S2R}.

 The factor $S_0(k)$ produces a background phase shift
 $\delta_{0}(k)=\frac{1}{2i}\log S_0(k)$
 (see e.g.  \cite{Taylor}).
  In the vicinity of the resonance energy  this
 is usually a slowly changing function of $k$ leading to a small shift of
 the resonance maximum.
 Therefore it is natural to expect that the cross section $\sg_h$ for the
 Hamiltonian $h$ will be close to the cross section $\sg_R$ \rf{sgR}.
 This property will be illustrated in the next section.

It is instructive to compare the scattering amplitude
 $f_{BW}(k)$ for the Breit-Wigner
$S$-matrix \rf{S2R} with the scattering amplitude $f_R(k)$ for $S_R$ given in \rf{Sh}.
From Eq. \rf{fk} one respectively finds
\[
\frac{1}{f_{BW}(k)}=\frac{b^2+(d+ik)^2}{-2d}
\]
and
\[
\frac{1}{f_R(k)}=\frac{1}{f_{BW}(k)}+\D(k)
\]
where
\be{it}
\D(k)=\frac{1}{|f_{BW}|}=
\frac{\bigl[(k^2+d^2-b^2)^2+4b^2d^2\bigr]^{1/2}}{-2d}
\ee
may be called the interference term.
From here according to Eq. \rf{gk},
 it follows that the effective range function $g_R(k)$ for $S_R(k)$
 is expressed in terms of
 Breit-Wigner effective range function
\[
g_{BW}(k)=\Re{\frac{1}{f_{BW}(k)}}=\frac{k^2-b^2-d^2}{2d}
\]
and the interference term as
\[
g_{R}(k)=g_{BW}(k)+\D(k)\,.
\]
In the next section using a simple toy model,
we will show that the interference term gives an essential contribution
to the cross section $\sg_R$ as compared to the cross section $\sg_{BW}$.
It not only disturbs the Lorentzian behavior of $\sg_{BW}$ curve but also
decreases twice the value of its maximum.

 \section{A toy model}

To illustrate general properties of the Hamiltonian $h$ we choose the
Hamiltonian $h_0$ with the
  simplest ($\nu=1$) scattering potential singular at the origin
\[
 v_0=2a_1^2\sinh^{-2}(a_1x)\,,\quad a_1\in\mathbb{R}\,,\quad a_1\ne0
 \]
 with the scattering states of the form
 \[
\psi_k(x)=\frac{\sqrt2}{\sqrt{\pi(k^2-a_1^2)}}
\left[a_1\coth(a_1x)\sin(kx)-k\cos(kx)\right].
 \]
Using the asymptotic behavior of this function
\[
\psi_k(x)\propto (ia_1-k)\exp(-ikx)-(ia_1+k)\exp(ikx)
\]
we find the
background scattering matrix
\[
S_0(k)=\frac{a_1-ik}{a_1+ik}
\]
as well as the cross section
\[
\sg_0=4\pi(k^2+a_1^2)^{-1}\,.
\]
The transformation function with the asymptotic behavior as given in
\rf{fisqrt} has the form
\[
u(x)=\exp(ax)\,[\,a_1\coth(a_1x)-a\,]\,, \quad a=d+ib\,,\ d<0\,.
\]
From Eqs. \rf{Vsusy} and \rf{L} we find the
  complex scattering potential
\be{V}
V(x)=\frac{2a_1^2\left(a^2-a_1^2\right)}{[a_1\cosh(a_1x)-a\sinh(a_1x)]^2}
\ee
(previously obtained in \cite{my-SS}) which, after a change of parameters, can
be reduced to a complex version of the well known one-soliton potential.

\begin{figure}[h]
\begin{center}
\epsfig{file=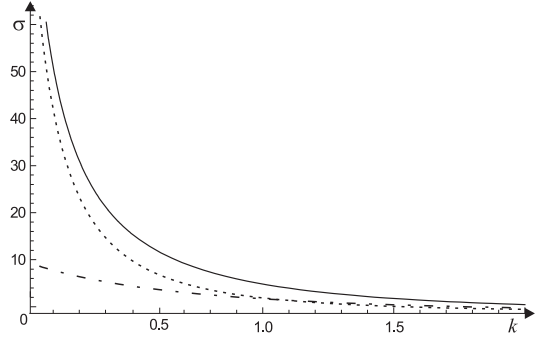, width=7.0cm}
\caption{
Scattering cross sections for the non-Hermitian Hamiltonian $H$:
 $\sg_e$ (the dot-dashed curve),
$\sg_r$ (the dotted curve) and $\sg_t$ (the solid curve).
} \label{fig3}
\end{center}
\end{figure}

Fig. \ref{fig3} shows
 the cross sections $\sg_e$ (the dash-dotted curve),
$\sg_r$ (the dotted curve) and $\sg_t$ (the solid curve) related to the scattering
matrix $S_H$.
(Everywhere the units are such that $\hbar^2/(2m)=1$ and $a_1=3$,
$b=0.5$).
\begin{figure}[h]
\begin{center}
\epsfig{file=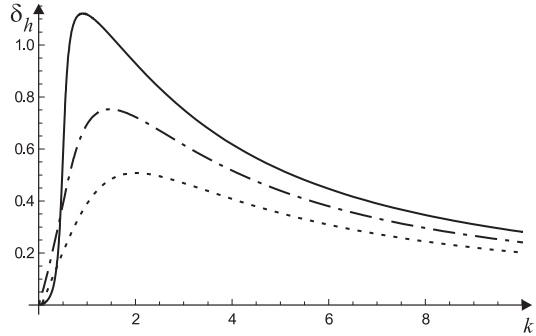, width=7cm}
\caption{Phase shifts $\d_h$ for
  $d=-0.1$ (the solid curve), $d=-0.5$ (the dot-dashed curve) and $d=-1.0$ (the dotted curve).
} \label{fig2}
\end{center}
\end{figure}
\begin{figure}[h]
\begin{center}
\epsfig{file=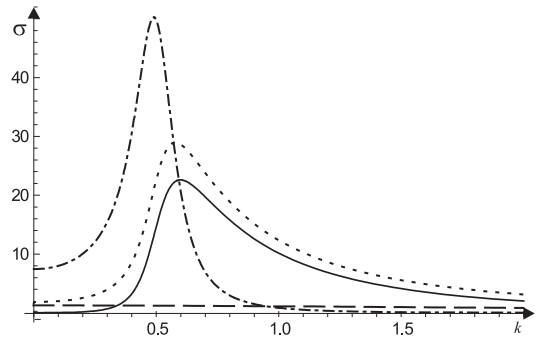, width=7cm}
\caption{Scattering cross sections $\sg_h$ (the solid curve),
 $\sg_R$ (the dotted curve), $\sg_{BW}$ (the dot-dashed curve) and $\sg_0$ (the dashed curve).
} \label{fig1}
\end{center}
\end{figure}
None of these curves exhibit any resonance behavior.
In contrast, for $d$ values close to zero,
 the scattering matrix $S_h$ gives rise to a pronounced maximum in the
 scattering cross section $\sg_h$.
Fig. \ref{fig2} illustrates the phase shift
$\d_h=\frac{1}{2i}\log S_h$ for $d=-1$ (the solid curve),
$d=-0.5$ (the dot-dashed
curve), and $d=-0.1$ (the dotted curve).
Near the point $k=0.5$ the smaller
the absolute value of $d$ is,
the  closer
  the corresponding
phase shifts come to the vertical line.

In Fig. \ref{fig1} we compare the behavior of different
cross sections for $d=-0.1$.
As already discussed, the background
cross section is close to zero for the whole drawn momentum interval,
(see the dashed curve).
The Hermitian Hamiltonian
$h$ cross section $\sg_h$ (the solid curve) is wider and less pronounced
 as compared to
the Breit-Wigner cross section $\sg_{BW}$ (the dot-dashed curve).
This agrees with the fact that $S_R$ is the square root of $S_{BW}$. The
difference between curves $\sg_R$ and $\sg_{BW}$ is due to the
interference term \rf{it} and it illustrates an important contribution
from this term to the cross section $\sg_R$.
The dotted and solid curves are close to each other.
This is because of a
small influence
that the background scattering matrix $S_0$ exerts on the resonance
scattering matrix $S_R$.

\section{Conclusion}

We have found a physical meaning for the phase factor of the
non-unitary scattering matrix $S_H$ of the non-Hermitian Hamiltonian $H$,
that is,
this phase factor is shown to be the same as the
unitary scattering matrix $S_h$
 of the Hermitian operator $h$ related to the given
non-Hermitian one by a similarity transformation.
We have demonstrated
that the possible presence of a spectral singularity in the continuous
spectrum of the non-Hermitian Hamiltonian $H$ translates as a resonance in the
scattering cross section of its Hermitian counterpart $h$
(see Fig. 3).
The closer in the space of parameters the Hamiltonian $H$ to
the singular point is,
the more pronounced becomes the resonance in the cross section of $h$.
This means that although the singular point is inaccessible for
the Hermitian Hamiltonian, it
 ``feels'' the presence of the singularity if it is ``close enough''.
Basing on this property one may conjecture
 that it might be possible to detect spectral
singularities in scattering experiments.
We explain the resonance behavior of the Hermitian Hamiltonian cross
section by the fact that corresponding scattering matrix up to a
background scattering matrix is a square root of the Breit-Wigner
scattering matrix.

\ack
The author is grateful to A. Pupasov for useful discussion.

\section*{References}


\end{document}